\documentclass[a4paper,twocolumn,11pt]{quantumarticle}
\pdfoutput=1
\usepackage[T1]{fontenc}
\usepackage[utf8]{inputenc}
\usepackage[numbers]{natbib}
\usepackage[colorlinks=true,citecolor=NavyBlue,linkcolor=RubineRed,urlcolor=Cerulean,pdfencoding=auto]{hyperref}
\usepackage{amsthm}

\usepackage[inkscapelatex=false]{svg}
\usepackage[final,commentmarkup=uwave]{changes}
\usepackage[]{todonotes}
\definechangesauthor[name={Benedikt}, color=purple]{BT}
\definechangesauthor[name={Anders}, color=green]{AS}
\definechangesauthor[name={Tzula}, color=orange]{TzBP}
\def\abs#1{\mathinner{|{#1}|}}
\def\abssq#1{\mathinner{|{#1}|}^2}

\usepackage{graphicx,amsmath,dsfont}
\usepackage[dvipsnames]{xcolor}
\usepackage{animate}
\usepackage{tabularx}
\usepackage{enumerate}
\usepackage{float}

\usepackage[capitalize]{cleveref}

\newcounter{protocol}

\usepackage{color}
\usepackage{amsthm}
\usepackage{amsmath,amssymb,amsbsy}

\usepackage{xcolor}
\pagecolor{white}
\usepackage[english]{babel}

\usepackage{calrsfs}
\newcommand{\lettersection}[1]{\section{#1}}

\newcommand{\had}{\hat{a}^\dagger}

\newcommand{\ha}{\hat{a}}

\newcommand{\vac}{|{\rm vac}\rangle}
\newcommand{\ket}[1]{\left| #1 \right\rangle}
\newcommand{\bra}[1]{\left\langle #1 \right|}
\newcommand{\braket}[1]{\left\langle #1 \right\rangle}

\newcommand{\proj}[1]{| #1\rangle\!\langle #1 |}

\newcommand{\eea}{\end{eqnarray}}
\newcommand{\bea}{\begin{eqnarray}}
\newcommand{\ee}{\end{equation}}
\newcommand{\be}{\begin{equation}}

\begin{document}

\title{Remotely Preparing Many Qubits with a Single Photon}

\author{Tzula B. Propp}\altaffiliation{These authors contributed equally and are corresponding authors: propp@physics.leidenuniv.nl; benedikt.tissot@nbi.ku.dk}
\affiliation{Leiden Institute of Physics\\ Leiden University, 2333 CA Leiden, The Netherlands}\affiliation{QuTech, Kavli Institude of Nanoscience\\
TUDelft, Lorentzweg 1, 2628 CJ Delft, Netherlands}
\author{Benedikt Tissot}\altaffiliation{These authors contributed equally and are corresponding authors: propp@physics.leidenuniv.nl; benedikt.tissot@nbi.ku.dk}
\affiliation{Center for Hybrid Quantum Networks (Hy-Q), Niels Bohr Institute, University of Copenhagen, Jagtvej 155A, Copenhagen DK-2200, Denmark}
\author{Anders S. Sørensen}
\affiliation{Center for Hybrid Quantum Networks (Hy-Q), Niels Bohr Institute, University of Copenhagen, Jagtvej 155A, Copenhagen DK-2200, Denmark}
\author{Stephanie D. C. Wehner}
\affiliation{QuTech, Kavli Institude of Nanoscience\\
TUDelft, Lorentzweg 1, 2628 CJ Delft, Netherlands}

\begin{abstract}
A single photon in a superposition of $d$ modes naturally encode a $d$-dimensional quantum system, a so-called qudit. We show that such superpositions can be leveraged to achieve a quantum speed-up of remote remote state preparation (RSP): a primitive for several quantum network protocols. 
For a superposition over $d\geq 2$ modes, the photon state can encode up to ${\rm Log}_2(d)$ qubits, which we exploit in a proposed reflection based RSP protocol with multiple variations. For single qubit RSP, we achieve a performance comparable to the best known existing schemes but with reduced requirements for phase stabilization. For many qubit RSP the achievable success rates remain high despite needing exponentially many temporal modes, since only one photon needs to be transmitted and detected to prepare multiple qubits. By simultaneously preparing many qubits at once, we bypass limited qubit lifetimes and improve fidelities beyond what is achievable with existing RSP protocols. 


\end{abstract}

\maketitle
\lettersection{Introduction}
Blind Quantum Computing (BQC) is a quantum internet meta-application \cite{Kimble2008,Wehner2018} enabling simultaneous quantum advantages in computational power \cite{Grover1997,Harrow2009,Mantri2013,BermejoVega2018,Seddon2021,Liu2021,Pokharel2023,Jordan2025,BeneWatts2025} and privacy \cite{Broadbent2009,Morimae2012,Mantri2017,Fitzsimons2017,Drmota2024}\footnote{There is also evidence for an advantage in energy consumption \cite{FellousAsiani2023,Meier2025,Hou2025,Abderrahim2025}, albeit not yet for networks.}, while concurrently allowing remote benchmarking of untrusted devices \cite{Liao2022,verification2024}. BQC implementations of aglorithms have grown in sophistication as the underlying technology matures \cite{Lanyon2013,Drmota2024,Ringbauer2025}. However, scalability is hampered by large resource demands of verifiable noise-robust BQC \cite{Takeuchi2019,Leichtle2021}, including the high demand for remote state preparation (RSP) of qubits \cite{Bennett2001,Leung2003}. The number of RSP qubits scales at least linearly with the number of logical qubits and the circuit depth \cite{Broadbent2009,Takeuchi2019}.
In the presence of decoherence and without quantum error correction it is vital to minimize the time spent in noisy storage \cite{Gorin2007,Brennen2008,Bumer2024}, e.g. for the first RSP qubit while the remaining RSP qubits are prepared.  Despite continued improvements in single-qubit RSP protocols \cite{Berry2003,Babichev2004,Barreiro2010,Tchebotareva2019,Alshowkan2021,vanDam2025,Propp2025}, low rates thwart scalability of qubit-by-qubit RSP in the era of memory-limited noisy intermediate-scale quantum (NISQ) devices \cite{vanDam2024}.

Verifiable BQC requires that a client use RSP to privately prepare many single-qubit states $\ket{+_{\theta}} = \left(\ket{0} + e^{i\theta} \ket{1}\right) / \sqrt{2}$ with the angles $\theta$ known only to the client. These phase-rotated qubits are then entangled into a brickwork graph or a more sophisticated non-planar BQC-compatible graph for a measurement based computation \cite{Raussendorf2001,Fitzsimons2017,Takeuchi2019b,Xu2020}. Verification is performed via computations with efficiently calculable, deterministic outcomes. 

An ideal $n$-qubit RSP protocol would simultaneously prepare an entire $n$-qubit product state 
\begin{align}
\label{eq:targetstate}
\ket{\Large {+}_{\vec{\theta}}} & \equiv \bigotimes_{i=0}^{n-1} \ket{+_{\theta_i}} \\
& = \frac{1}{\sqrt{2^n}} \sum_{\vec{x}=\{0,1\}^n} \exp \left( {i\sum_{l=0}^{n-1} \theta_l x_l} \right) \ket{\vec{x}} \notag
\end{align}


Beyond BQC, the ideal many-qubit RSP in Eq.~\eqref{eq:targetstate} is beneficial for e.g. quantum key distribution over repeater chains \cite{vanDam2025}, quantum network initialization and benchmarking \cite{Chen2024}, error-corrected delegated quantum computation \footnote{Many-qubit RSP is less resource demanding than e.g. logical qubit teleportation \cite{Fowler2010,Luo2021,Eckstein2024}, and since physical qubits are intrinsically susceptible to decoherence e.g. in a surface code \cite{Poulin2005,Bacon2006,Raussendorf2007,Fowler2012,Zhao2022,Sivak2023,google2024}, one would ideally like to prepare them simultaneously. For the role of RSP in delegated computation, see Ref. \cite{Chen2024}.}, with straightforward generalizations for many-party protocols e.g. multi-client BQC, secret sharing, and quantum position verification \cite{Polacchi2023,Cleve1999,Kanneworff2025,Polacchi2025}.

\subsection{Contributions}
In this article, we introduce reflection based RSP (R-RSP), which can deterministically produce the state $\ket{\Large {+}_{\vec{\theta}}}$ in Eq. \eqref{eq:targetstate} with a high fidelity heralded on detection of a single photon in superposition over many modes. Because the preparation of the entire n-qubit state is heralded simultaneously  by detection of a single photon, the protocol provides a significant advantage in situations where multiple qubits are required simultaneously; it avoids memory decoherence occurring between sequential events in individual RSP.  This can provide a significant advantage for long-distance multi-qubit RSP. We introduce $8$ variations of our protocol defined by 3 choices: single or many qubits, single photons or weak coherent pulses (WCPs), and client devices that emit or detect light. The underlying principle can thus be exploited in multiple setting depending on the available hardware. 

Our R-RSP protocol combines several benefits of existing RSP protocols: it is fast like emission based single-click RSP \cite{vanDam2025}, amenable to a variation with WCPs like emission based single/double/double-single-click RSP \cite{Babichev2004,Peters2005,Rosenfeld2007,Iuliano2024,vanDam2025}, relaxes the need for phase stabilization like double/double-single-click RSP \cite{vanDam2025}, and amenable to a measurement based variation \cite{Drmota2024}, where security is guaranteed by quantum state teleportation and the no-cloning theorem \cite{Wootters1982,Bennett1993,Barnum1996}.
To the authors' knowledge the proposed protocol is the \emph{first} to remotely prepare multiple qubits in stored memory upon the detection of a single photon.

\lettersection{Protocol} 
The key resource in our protocol is a light-matter state
\begin{align}
\label{eq:resourcestate}
\ket{\Phi} =  \sum_{\vec{x}=\{0,1\}^n} c_x \hat{a}_x^\dagger\vac \ket{\vec{x}} ,
\end{align} with $\hat{a}_x^\dagger$ the bosonic creation operator for the $x$th (temporal) mode, satisfying $[\ha_x,\had_{x'}]=\delta_{x,x'}$.
As we show below, a state of this form can be used to remotely prepare the $n$-qubit state $\ket{\Large {+}_{\vec{\theta}}}$---up to single-qubit corrections---by detection of a single photon \footnote{For verification rounds, one can remotely prepare ``dummy'' states $\ket{0/1}$ for to de-entangle neighboring qubits \cite{Fitzsimons2017b,Leichtle2021,Yang2025}. In our protocol, dummy states can be prepared by setting the coefficients $c_x$ to zero for certain bitstrings $\vec{x}$. While this results in information about the number of dummy states being revealed by photon transmission statstics, BQC protocols with a revealed number of dummies/traps per layer can be secure \cite{Kashefi2017}.  However, dummy states are not strictly necessary \cite{Hayashi2015,Fitzsimons2017}; stabilizer state preparation can be used instead.}.

\begin{figure}[t]
\centering
\includegraphics[width=\linewidth]{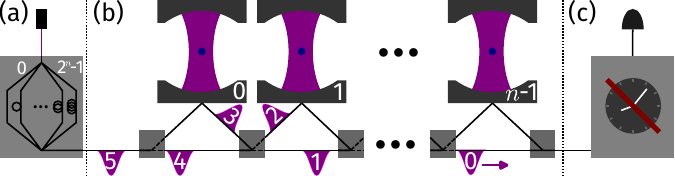}
\caption{\label{fig:proto}Sketch of the setup for generalized remote state preparation of a multi-qubit register of length $n$.
The steps are (a) preparation of a train of $2^n$ time-bins (purple), (b) the qubit-photon interaction, and (c) the removal of the which time-bin information (e.g. using a quantum Fourier transform or time-lens \cite{Barak2007,Donohue2016}) followed by single click heralding.
The server does two of the steps including (b) and the remaining step, either (a) or (c), is performed by the client.
The client additionally imprints individual phases on the time-bins, either after splitting in (a) or before removing the which-time-bin information in (c).
The qubit photon interaction (b) is engineered by
optical switches (small gray boxes) controling the path of the time-bins and thus with which qubits a time-bin interacts.
The protocols requires that time-bin $x$ with binary representation $\vec{x} \in \{0,1\}^n$ interacts with qubit $l$ implementing a conditional phase flip iff $x_l = 1$.}
\end{figure}

It is possible to generate the state \eqref{eq:resourcestate} by engineering the light-matter interaction,
and we present an approach inspired by reflection based entanglement generation protocols \cite{nemoto-2014-photon_archit_scalab_quant_infor_proces_diamon,Omlor2025,Singh2025}, quantum computing enhanced optical imaging \cite{mokeev-2025-enhan_optic_imagin_via_quant_comput}, and high-dimensional entanglement generation \cite{Zheng2022,Bharos2025}. 
In the following we discuss an implementation using a conditional phase flip \cite{duan-2004-scalab_photon_quant_comput_throug} that can be implemented using efficient matter-photon interfaces.
Such efficient matter-photon interfaces have been demonstrated in many systems, e.g.,
trapped atoms \cite{Reiserer2014}, 
quantum dots \cite{sun-2016-quant_phase_switc_between_singl}, 
and silicon vacancy color centers in diamond \cite{nguyen-2019-integ_nanop_quant_regis_based,nguyen-2019-quant_networ_nodes_based_diamon}.

In the first step of the protocol, a photon in a superposition over time-bins $\ket{\psi}=\sum_x c_x\hat{a}_x^\dagger\vac$ is generated, see Fig.~\ref{fig:proto}(a), while the server prepares a tensor product state $\ket{+}^{\otimes n}$, e.g., by preparing \(\ket{0}^{\otimes n}\) and applying a single-qubit Hadamard gate \(\mathbf{H} = \frac{1}{\sqrt{2}}\begin{pmatrix} 1 & 1 \\ 1 & -1 \end{pmatrix}\) on each qubit of the register. We implement a gate $U_{q,x}$ between qubit $q$ and the photon wave function component in the time-bin $x$ by routing the time-bins such that a photon in time-bin $x$ interacts with qubit $q$ only if the $q$'th binary digit of $x_q = 1$, as illustrated in Fig. \ref{fig:proto}(b); thus, if a photon is present in mode $x$, each qubit $q$ in the register corresponding to a $1$ in bitstring $\vec{x}$ transforms $\ket{+}\rightarrow \ket{-}$. In the ideal case, this enables implementation of the multi-qubit multi-mode entangling gate $U$:


\begin{align}
 \label{eq:CPHASE} & U=\otimes_{x=0}^{2^n -1}\otimes_{q=0}^{n-1} U_{q,x}, \text{ with } \\
  \label{eq:CPHASEqx}
  & U_{q,x}:\,\,\ket{k}_q (a_x^{\dag})^l \vac \to (-1)^{klx_q} \ket{k}_q (a_x^{\dag})^l \vac . 
\end{align}
By applying the multi-qubit multi-time-bin  gate \(U\) followed by a Hadamard $\mathbf{H}$ on each qubit, we generate the light-matter state in Eq.~\eqref{eq:resourcestate}. Then which-time-bin information is erased e.g., via a quantum Fourier transform (QFT) or time lens \cite{Barak2007,Donohue2016}.
Provided the \emph{posteriori} probability that a photon was in each mode is identical, detection of a single photon [Fig.~\ref{fig:proto}(c)] then heralds the desired RSP state in Eq.~\eqref{eq:targetstate} up to single-qubit phase corrections, as detailed in Appendix \ref{app:imperfections}. 
While here we assume each qubit interacts with the optical field via a reflection and also name the protocol accordingly, it is also possible to implement this protocol using only a single communication (Raman) qubit that is fully connected to the entire $n$-qubit register (and an additional parity-check qubit if absorption is used to map the photon state to the interface qubit instead of reflection), see Appendix \ref{app:absorb}. 


To use the reflection based gate \eqref{eq:CPHASE} to remotely prepare the $n$-qubit state in Eq. \eqref{eq:targetstate} with high-fidelity, it is sufficient for the client to either generate the state $\ket{\psi_c}= \sum_x c_x \hat{a}_x^\dagger\vac $, or perform a measurement of the photonic component of the resource state in Eq.~\eqref{eq:resourcestate} that projects onto a set of states $\ket{\chi}_{\ell}=\sum_x d_{x,\ell} \hat{a}_x^\dagger\vac $.
In either case, the phases of $c_x$ ($d_{x,\ell}$) can be chosen by the client to encode random phases on the qubits (up to local corrections), and the amplitudes of $c_x$ ($d_{x,\ell}$) can be chosen by the client to compensate for unequal losses.



As we will show in the following, multi-qubit R-RSP features favorable scaling with fiber losses -- which is expected to be the dominant source of losses for long-distance RSP. We achieve a total success probability $P_0$ to prepare all $n$ qubits that is linear in the transmission efficiency. However, the need for an exponential number of temporal modes requires a longer attempt duration, inducing a rate-optimization problem that we explore at the end of this article. 
\subsection{R-RSP Ideal Implementation}
We focus here on the emission based clients as they can continuously attempt until success, decoupling the trial time from the communication time, which is a major advantage for long distances.
In this variation of the protocol, the client prepares a photon distributed over $2^n$ time-bins, and encodes the target phases in the phase-differences of the time-bins.
Given the implementation illustrated in Fig. \ref{fig:proto}, we introduce efficiencies \(\eta_t, \eta_0, \eta_1,\) and \(\eta_d\). They correspond respectively to the probabilities to transfer a single photon from the client to the register, through the non-interacting branch, the interacting branch, and from the register up to and including detection. Note that \(\eta_d\) includes the efficiency of the removal of the time-bin information and \(\eta_0, \eta_1\) include the efficiency of the photon routing. Finally, \(1-\eta_t\) unites emission inefficiencies and fiber losses such that we expect $\eta_t$ to be the lowest efficiency.
With these, we can specify the transmission efficiency $\eta_x$ for each mode: $\eta_x = \eta_t \eta_d \eta_1^{H(x)} \eta_0^{n-H(x)}$, with $H(x)$ the Hamming weight (number of $1$'s) of $\vec{x}$.

Accounting for the efficiencies $\eta_x$, the overall probability of detection is \(P_{0} = \sum_{\vec{x} \in \{0,1\}^n} |c_x|^2 \eta_x\).
Choosing amplitudes satisfying \(\abs{c_x}^{-2} = \eta_x {\sum_{\vec{y} \in \{0,1\}^n} \eta_y^{-1}}\),
ensures the probability for the photon coming from any mode to be identical
and thus resulting in the correct target-state conditioned on detection of a pulse.
We arrive at a single photon detection probability
\begin{align}\label{eq:detectionprobfinal}
P_{0} = \eta_{t} \eta_{d} \left( \frac{2 \eta_0 \eta_1}{\eta_0 + \eta_1} \right)^n .
\end{align}
We highlight that the detection probability scales linearly with the efficiency of fiber transmission $\eta_t$, since only one photon in total needs to be transmitted; practically, the tradeoff is the need to generate and manipulate a superposition of a single photon over $2^n$ modes. 
As we will show below, this aids in resolving the windowing problem of NISQ devices (i.e. the problem of preparing multiple qubits with the decoherence time), as the rate to prepare all \(n\) qubits retains a linear scaling in fiber transmission.
By choosing phases satisfying \(  \phi_x = \sum_{l=0}^{n-1} \theta_l x_l ,\) the resulting state is precisely that in Eq. \eqref{eq:targetstate} with (random) phases $\theta_l$ applied to each qubit. 
Since there are $2^n$ phases $\phi_p$ in the photonic superposition and only $n$ target qubit phases $\theta_l$, it is always possible to satisfy this constraint. 


\lettersection{R-RSP with Imperfections} Having discussed the influence of photon loss, we now include errors due to imperfect state preparation at the client side.
The imperfection we account for are
modes not perfectly compatible with the server and two-photon errors when using weak coherent pulses \cite{Dunjko2012,Jiang2019}.
In Appendix \ref{app:imperfections} we show how to model the protocol in the presence of these errors.
We parametrize both errors in terms of the client state, by introducing amplitudes $\alpha_k$ for the $k$ photon component of the wave function
and splitting the single photon component into a mode that perfectly adheres to the gate $\propto 1-\epsilon$ and an orthogonal mode that does not $\propto \epsilon$.
By assigning zero fidelity to the part of the state that upon impinging on the register either is in a multi-photon or the orthogonal mode, we find a bound for the fidelity
\begin{align}
    \label{eq:fidelity_main}
    & F \ge (1-\epsilon)  \frac{P_0}{P} \left[ \abssq{\alpha_1} + 2 \abssq{\alpha_2} (1-\eta_t) \right] \\
    \,& \approx ( 1 - \epsilon ) \left\{ 1 - \frac{2 \abssq{\alpha_2}}{\abssq{\alpha_1}} P_0 \left[ \frac{1}{\eta_d} \left( \frac{\eta_1 + \eta_0}{2 \eta_0 \eta_1} \right)^n - 1 \right] \right\} ,
    \notag
\end{align}
with the success probability when using a photon number resolving detector given by
\begin{align}
  \label{eq:detectionprob_twophoton}
    P = \abssq{\alpha_1} P_0 + 2 \abssq{\alpha_2} P_0 \left( 1 - P_0 \right) ,
\end{align}
in terms of the single-photon success probability \(P_0\) according to Eq.~\eqref{eq:detectionprobfinal}.


\begin{figure}[t]
\centering
\includegraphics[width=\linewidth]{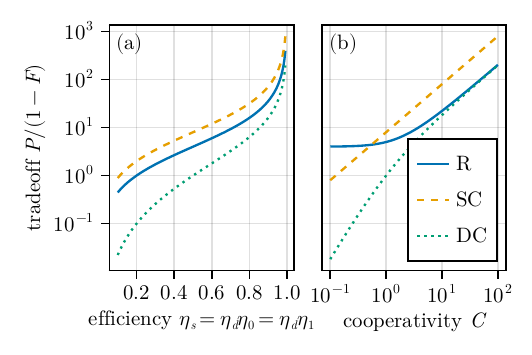}
\caption{\label{fig:trade-off-n1} Tradeoff between fidelity $F$ and success probability $P$ when using a weak coherent pulse and photon number resolving detection. We compare the reflection based (R), single-click (SC), and double-click (DC) remote state preparation protocols, see legend.
(a) a scenario limited by photon routing and detection where we take $\eta_s = \eta_0 \eta_d = \eta_1 \eta_d$ and
(b) a scenario limited by the matter-photon interaction where we consider $\eta_0,\eta_d=1$, $\eta_s = C/(1+C)$, and $\eta_1 = [C/(2+C)]^2$.
Note that we do not account for the contribution of phase fluctuations which are expected to have the strongest detrimental effect for the SC protocol.
}
\end{figure}

\lettersection{Results: Single-Qubit R-RSP} Having introduced the reflection based protocol, we begin by comparing it to single qubit preparation protocols.
Using our error model in Eq.~\eqref{eq:fidelity_main} and assigning a zero-fidelity contribution to multi-photon and orthogonal mode components, we can also calculate the average fidelity due to the remaining two-photon components (see Appendix \ref{app:imperfections}). 
We find the leading order rate-fidelity tradeoff for $n=1$ and $\epsilon=0$ 
\begin{align}
\label{eq:rate-fidelity-n=1}
    F \approx 1 - \frac{P}{4} \left( \frac{1}{\eta_0 \eta_d} - \frac{2 \eta_1}{\eta_0 + \eta_1} \right) .
\end{align}
We focus on emission based protocols, as they decouple the success probability and transmission time, and reserve discussion of measurement protocols for Appendix \ref{app:R-R-RSP}. 
The idea of those protocols is that the server qubits emit (part of) a photon
which is combined with the client signal on a beamsplitter, such that the detection of a photon in the combined channels heralds the transfer of a phase to the server qubit.
Possible implementations are 
 the single (SC) and double-click (DC) protocols \cite{vanDam2025},
respectively using a single or two clicks to herald the transfer of the phase.
We emphasize that these protocols have not been shown to be extendable to prepare $n>1$ qubits. It is thus unclear if they can attain the the advantages of the $n>1$ protocol developed here,
Furthermore, SC-RSP has stringent phase stability requirements \cite{vanDam2025}.

In a regime where the emission intensity of the client is limited by the fidelity of the state transfer, a good merit of comparison is the tradeoff between success probability (or rate) and fidelity.
The rate-fidelity tradeoffs for photon number resolving detectors of these protocols \cite{vanDam2025} reads
$F_{SC} = 1 - \frac{P_{SC}}{8} \frac{1-\eta_s}{\eta_s}$ and
$F_{DC} = 1 - \frac{P_{DC}}{2} \frac{1-\eta_s}{\eta_s^2}$.
These only depend on the server efficiency $\eta_s$, which describes the probability that a photon emitted from the server qubit leads to a click in the detector.
Here and in Eq.~\eqref{eq:rate-fidelity-n=1} we see that $P_{p}/(1-F_p)$ corresponds to a protocol $p$ and system parameter dependent constant quantifying the performance.
To compare the tradeoffs we consider two regimes. The first one is the regime where the qubit-photon interface is efficient, such that the inefficiency is dominated by the photon routing and detection.
In this scenario we consider $\eta = \eta_0 \eta_d = \eta_1 \eta_d = \eta_s$ and we find $\frac{P_R}{1-F_R} = \frac{1}{2} \frac{P_{SC}}{1-F_{SC}} > \frac{P_{DC}}{1-F_{DC}}$, i.e., SC-RSP achieves twice the rate of the reflection ($R$) based protocol both of which outperform DC-RSP.
In Fig.~\ref{fig:trade-off-n1}(a) we display the tradeoff $P/(1-F)$
confirming that in this regime the reflection based scheme is close to the SC scheme in performance, but with greatly relaxed phase stability requirements.
Furthermore, the $SC$ and $R$ schemes greatly exceed the $DC$ scheme for low efficiencies.
For $\eta \to 1$ all tradeoffs diverge if single photon detectors are used due to the fidelity reaching unity even for a non-vanishing P.

The second regime we consider is if the routing and detection are very efficient and the bottleneck is the qubit-photon interface [Fig.~\ref{fig:trade-off-n1}(b)]. 
To compare this regime we express the efficiencies of the protocols in terms of the cooperativity $C$ of a cavity quantum electrodynamics system.
For the emission based protocols we use an established upper bound for the cavity enhanced photon retrieval efficiency $\eta_s = C/(1+C)$ \cite{gorshkov-2007-photon_storag_in,tissot-2024-effic_high_fidel_flyin_qubit_shapin,kollath-boenig-2024-fast_storag_photon_cavit_assis_quant_memor}
while for the reflection based protocol we use the interaction efficiency $\eta_1 = [C/(2+C)]^2$ (see Appendix \ref{app:spinphoton}) 
and take the remaining efficiencies to be $\eta_d=\eta_0=1$.
The square difference between the efficiencies can be understood as follows: in emission the photon only needs to exit the cavity, but for conditional reflection it needs to enter and exit the cavity.
In this regime the R-RSP tradeoff exceeds both SC-RSP and DC-RSP in the low cooperativity limit; this limit also places stringent requirements on the pulses that can be used from the sending side, as the reflected and transmitted wavepackets need to be indistinguishable (see 
Appendix \ref{app:intensityRSP}
for an alternative qubit-photon interaction mitigating this).
In the large-cooperatively limit R-RSP approaches the performance of the DC scheme, while the SC scheme again performs best but requires phase stabilization over the long-distance fiber \cite{vanDam2025}.

\added[id=BT]{If the photon intensity emitted by the sender is not limited by the state fidelity but other constraints, 
instead of the tradeoff, the success probabilities should be compared directly. 
The success probability for R-RSP \eqref{eq:heralding-probability} to leading order yields $\eta_t \eta_d \frac{2 \eta_0 \eta_1}{\eta_0 + \eta_1} \abssq{\alpha_1}$.
In comparison, $P_{DC} \sim \frac{\eta_t \eta_d \eta_s}{2} \abssq{\alpha_1}$ and $P_{SC} \sim 2 \eta_t \eta_d \abssq{\alpha_1}$.}




\begin{figure}[t]
\centering
\includegraphics[width=\linewidth]{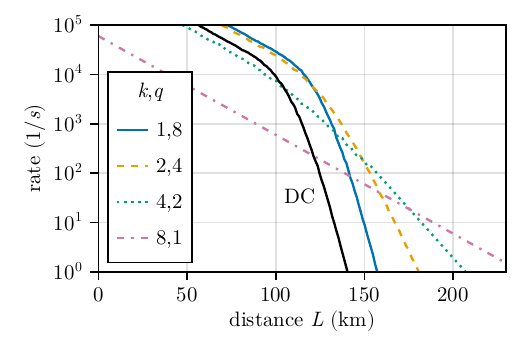}
\caption{\label{fig:8qubits}
Rate to prepare $n=k q =8$ qubits within a sliding window of $w=2000/2^{k-1}$ attempts using single-photon sources as a function of distance.
Different distribution in $q$ batches (photon detections) and $k$ qubits prepared using a single photon are encoded in the linestyle, see legend. We also include the performance of DC-RSP.
We assume $\eta_t$ corresponding to transmission over a distance $L$ in a fiber with a loss rate of $0.2\,\text{db}/\text{km}$ and the remaining quantum efficiencies during emission are in $\eta_{t,\text{intrinsic}} = 0.9$.
Additionally we assume efficient spin-photon interaction with $\eta_0 = 0.9$ and $\eta_1 = \eta_0 [C/(C+2)]^2$ with $C=38$ as well as a detection efficiency of $\eta_d = 0.9$. We take the pulse duration to be $T_{\text{TB}} = 30\,$ns.}
\end{figure}

\lettersection{Results: Many-Qubit RSP} We now turn our attention to remotely preparing many qubits with a single photon.
This allows reflection based protocols to simultaneously prepare many high-fidelity qubit states. 


Increasing the number of qubits a single photon prepares in the many-qubit R-RSP protocol also exponentially increases the number of time bins needed \footnote{The large number of time bins needed can be partially mitigated by spatial mode multiplexing (e.g. pairing pulses corresponding to bitstrings $\vec{x}$ and $\bar{\vec{x}}$ which will not interact with the same qubit registers to reduce the needed interaction time by a factor of $2$), using other photonic encodings, or engineering multi-qubit multi-mode light-matter interaction \cite{Higgins2014,Kagalwala2017,Bosman2017,Garziano2020}.}. Thus, there is an optimal RSP batch size $1 \le k \le n$ depending on the efficiencies, qubit number, qubit coherence time, target fidelity, and the time $T_{\text{TB}}$ each additional time-bin adds to the total pulse train duration $(2^{n}-1) T_{\text{TB}}$. To illustrate this we cast it into a situation where a number of attempts may be made before discarding a decohered prepared qubit, i.e., the ``window problem'' of scan statistics \cite{Davies2024}.
If we require that all qubits are prepared within the time it takes to make $w_0$ single qubit RSP attempts, each $k$-qubit attempts takes  $2^{k-1}$ times the single-qubit duration, allowing $w=w_0/2^{k-1}$ attempts.
We consider preparing $q$ batches of $k$ qubits to prepare all $n=qk$ qubits within a window of $w$ trials. Numerically, we simulate trajectories to calculate the average number of trials until success and estimate the rate, with the full implementation available in Ref.~\cite{CODE}.

For $w_0=2000$ we show numerical results in Fig.~\ref{fig:8qubits}.
As the distance increases, we observe in Fig.~\ref{fig:8qubits} a rapid decrease of the rate of the communication for low $k$  as the protocol is limited by the coherence window. This is in contrast to the high $k$ behavior  which  show a slower decrease since fewer photons need to be received within the window. This results in a many order of magnitude improvement in rates. This underlines the large performance benefit quantum multiplexing can achieve \cite{LoPiparo2019,Propp2025};
coherence between pulses yields a beyond-classical advantage to the rate, here from a qudit encoding that would not be possible using incoherent multiplexing techniques. 

This many orders of magnitude improvement can also be understood as a transition to the asymptotic behavior described analytically in Ref.~\cite{Davies2024}, where $R_q \propto \eta_t^q$ (see also 
Appendix \ref{app:asymptot}).
Emphatically, the total success probability decreases when increasing $n$ for a fixed window size $w_0$; therefore, we expect to reach this asymptotic scaling at shorter distances as the number of qubits $n$ increases.

In the absence of a window (i.e., $w_0=\infty$) an additional effect we want to highlight can be seen in the fidelity bound \eqref{eq:fidelity_main}
by only focusing on the intrinsic infidelity of the matter-photon interface $\epsilon$.
Considering the preparation of $n = q k$ qubits in a set of $q$ batches of $k$ qubits results in an $n$-qubit register fidelity $(1-\epsilon)^q \sim 1 - q \epsilon$;
 in terms of intrinsic errors, preparing multiple qubits using a single photon may give better performance. 




%

\lettersection{Conclusion \& Outlook} In this article, we have introduced a new reflection based RSP protocol demonstrating rate-fidelity tradeoff with weak coherent pulses outperforming DC-RSP protocol while having comparable phase stability requirements.
In some parameter regions it achieves performance comparable or exceeding SC-RSP with reduced need for phase stabilization.

In the NISQ era with modest memory lifetimes and in the absence of error correction, many-qubit R-RSP partially circumvents the fidelity decrease due to decoherence (or the rate decrease due to use of a cutoff window) through near-simultaneous preparation of many-qubit states. A security proof extending RSP to qudits is still needed \cite{Garnier2024} as are implementation-specific simulations \cite{coopmans2021netsquid}, but a new path forward to many-qubit  BQC in the NISQ era is now possible. There is, however, no free lunch; an exponentially large number of time bins are required, creating an R-RSP batch size optimization problem. In the high photon loss limit, we find it is fastest to remotely prepare many qubits with a single photon.

\lettersection{Acknowledgements} Our method is inspired by the ``bad wine'' riddle, where a regent knows that one of $1000$ wine bottles is poisoned with a slow acting poison but only has $10$ tasters available to find the poisonous bottle within the incubation period of the poison.

We also thank Janice van Dam, Emil R. Hellebeck, and Fabian Wiesner for useful conversations about this project. 

  This work was funded by the European Union's Horizon Europe research and innovation programme under grant agreement No.~101102140 – QIA Phase 1.
  BT and ASS acknowledge the support of Danmarks Grundforskningsfond (DNRF Grant No.~139, Hy-Q Center for Hybrid Quantum Networks). TzBP gratefully acknowledges support from the Quantum Software Consortium Ada Lovelace
Fellowship.

\bibliographystyle{quantum}
\bibliography{NTuple}

\appendix

\section{Alternative Implementation with a Single Communication Qubit}\label{app:absorb}\label{tempSM}
\begin{figure}[ht]
\centering
\includegraphics[width=\linewidth]{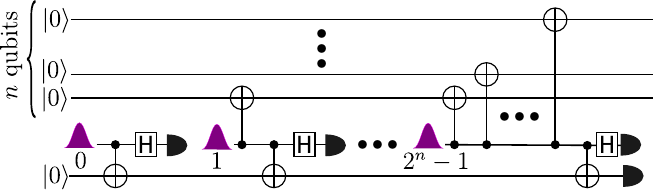}
\caption{\label{fig:ARSP}Circuit illustrating an alternative implementation of the protocol based on photon absorption.}
\end{figure}
Instead of implementing the state transfer and interaction with the register optically, we can also mediate it with a single interaction qubit and a heralding qubit.
In Fig.~\ref{fig:ARSP} we display a circuit showcasing how a communication qubit absorbing the time-bins combined with a heralding qubit can be used to implement the photon register interaction.
To understand the idea we follow the evolution of the system given a single photon input state
\(\ket{\Phi_p} = \sum_{x=0}^{2^n-1} c_x a^{\dag}_x \vac\).
The communication qubit is designed to absorb the time-bins \((\alpha + \beta a_x^{\dag}) \vac \ket{0} \to \vac (\alpha \ket{0} + \beta \ket{1})\) for any time-bin \(x\), this interaction can be implemented using a $\Lambda$-scheme \cite{gorshkov-2007-photon_storag_in,tissot-2024-effic_high_fidel_flyin_qubit_shapin,kollath-boenig-2024-fast_storag_photon_cavit_assis_quant_memor}.
Iteratively absorbing each time-bin in this manner, and transfering the corresponding amplitude of the time-bin to the heralding bit and the register qubits that have a \(1\) in the binary representation of \(\vec{x}\), see Fig.~\ref{fig:ARSP}.
After application of the CNOT gates, the communication qubit is reset by performing a Hadamard gate followed by readout and reset to \(\ket{0}\) to receive the next time-bin.
After receiving all time-bins we find the state transfer leads to
\begin{align}
  & \ket{\Phi_p} \ket{0}_c \ket{0}_h \ket{\vec{0}} = (\sum_{x=0}^{2^n-1} c_x a^{\dag}_x \vac) \ket{0}_c \ket{0}_h \ket{\vec{0}} \notag \\
  & \to 
  \label{eq:combined_state_single_comm}
  (\sum_{x=0}^{2^n-1} \sigma_x c_x \ket{\vec{x}}) \vac \ket{0}_c \ket{1}_h 
\end{align}
with \(\sigma_x = \pm\) corresponding to the communication bit being read-out in \(\ket{0}\) or \(\ket{1}\) after receiving the \(x\)'th time-bin.
The phases \(\sigma_x\) are known after the receiving all qubits and can thus be corrected. Since the interaction is absorption-based instead of reflection-based, there is also no need to erase which-time-bin information of the photon; its presence is confirmed by measuring the heralding qubit's state. 
Thus, we demonstrated that receiving (and heralding on the presence of the photon) also works with a single communication qubit.

\section{The Spin-Photon Gate}\label{app:spinphoton}

Central to our results is the ability to engineer a many-mode many-qubit interaction to produce the resource light-matter state in Eq.~\eqref{eq:resourcestate}. Here we describe in detail the cavity-QED interaction that enables this.

\begin{figure}[ht]
\centering
\includegraphics[width=0.6\linewidth]{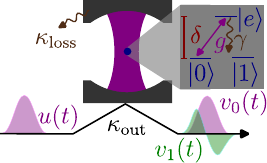}
\caption{\label{fig:ion-photon}Sketch of the input-output model for a single photon pulse interaction with a trapped ion mediated by a cavity.}
\end{figure}
We consider a system that can be described by a transition \(\ket{0} \leftrightarrow \ket{e}\) coupling to a cavity and if the qubit is in state \(\ket{1}\) it is decoupled from the cavity, see Fig.~\ref{fig:ion-photon}.
In an appropriate rotating frame this is described by the Hamiltonian
\begin{align*}
  H_s/\hbar = \delta \proj{e} + g (\ket{e}\bra{0} \hat{c} + \mathrm{H.c.}) ,
\end{align*}
where \(\delta\) is the detuning between the atomic transition and the cavity resonance frequency and $g$ is the coupling strength between the atomic transition and the cavity.
Considering the interaction with the photon channel with strength \(\kappa_{\text{out}}\) and accounting for orthogonal cavity losses at a rate \(\kappa_{\mathrm{loss}} = \kappa - \kappa_{\text{out}}\) as well as atomic losses at a rate \(\gamma\),
one can derive the conditioned transfer functions, e.g., following Ref.~\cite{Omlor2025} where an approach for input-output theory for quantum pulses \cite{Kiilerich2019} was used.
The results can be described in terms of transfer functions $r_k(\omega)$ that relate efficiency, input and output modes conditional on the qubit state $\ket{k}_q$ according to
 $\sqrt{\tilde{\eta}_k} v_k(\omega) = r_k(\omega) u(\omega)$.
Here $v_k,u$ describe normalized (but not orthogonal modes) and $\tilde{\eta}_k$ the reflection efficiency conditional on the state.
The transfer functions are
\begin{align*}
  r_{0}(\omega) &= 1 - \frac{2 \frac{\kappa_{\text{out}}}{\kappa} \left( 1-2i \frac{\omega-\delta}{\gamma} \right)}{\left( 1-2i \frac{\omega-\delta}{\gamma} \right) \left( 1 - 2 i \frac{\omega}{\kappa} \right) + C} \\
  r_{1}(\omega) &= 1 - \frac{2 \frac{\kappa_{\text{out}}}{\kappa}}{\left( 1 - 2 i \frac{\omega}{\kappa} \right)}
\end{align*}
with the cooperativity \(C = \frac{4 g^2}{\kappa \gamma}\)
and the detuning between the photon pulse frequencies and the cavity frequency \(\omega\).
For phase encoding the goal is \(r_0(\omega) \approx - r_1(\omega)\) for all relevant frequencies \(\omega\) of the input pulse. 
For sufficiently narrow band pulses choosing \(\frac{\kappa_{\text{out}}}{\kappa} = \frac{C+1}{C+2}\) leads to the reflection functions at resonance (\(\omega=\delta=0\))
\begin{align}
  \label{eq:refl_res}
  r_k(0) = (-1)^k \frac{C}{C+2} ,
\end{align}
as desired.
While we assumed perfect resonance for simplicity we note that the cavity system can be engineered to minimize finite bandwidth effects \cite{Omlor2025}.

In order to achieve a low error rate we assume the server (cQED system) is tailored towards the conditional phase using the choice above and the pulses are sufficiently narrow.
Thus resulting in the efficiency $\eta_1 = [C/(C+2)]^2$ used in the main text.
Additionally we assume that the sender either sends single photons or weak coherent pulses with a low probability of two photons.
Inspired by the above model we thus assume that the main part of the interaction implements the conditional phase evolution \eqref{eq:CPHASEqx}.
Additionally we split the single photon mode into a part \(u_0\) that performs the controlled phase flip gate in Eq.~\eqref{eq:CPHASEqx}, and a part \(u_{\perp}\) that does not perform the gate.
While here we do not differentiate between temporal and other kinds of modes, 
we take these modes to reflect the effect of any type of modes (including e.g., spatial).
Thus, we assume that the client's transmitted state can be described as an arbitrary superposition of vacuum and single/two-photon Fock states
\(\alpha_0 \ket{0}_0 + \alpha_1 (\sqrt{1-\epsilon} \ket{1}_0 + \sqrt{\epsilon} \ket{1}_{\perp}) + \alpha_2 \ket{2}_0\), assuming any higher-order contributions are negligible.
Here we consider both \(\abs{\alpha_2}, \epsilon \ll 1\), and \(\ket{k}_0\) and \(\ket{k}_{\perp}\) denote the Fock states of the photon in mode \(u_{0}\) and \(u_{\perp}\) respectively.
By definition \(\ket{k}_0\) satisfies \eqref{eq:CPHASEqx} for \(k=0,1\) and we also assume it to hold for \(k=2\).
The argument for this simplification is that, for sufficiently narrow band pulses, it is unlikely that two photons of a pulse interact with the system simultaneously. 
Thus the two photon transfer function can be approximated by the square of the single photon transfer function.
Instead of the actual effect of the scattering of \(\ket{1}_{\perp}\), for simplicity of the error analysis we assume that it simply does not lead to a phase difference at all, effectively encodes the correct single photon gate infidelity, and the temporal mode remains unchanged.
This is an assumption, made for simplicity, which yields simple lower bounds to the final fidelity. For simplicity we also disregard the effect of a partial phase imprint if a photon is lost during the process, such that we can include photon loss in the model by addng loss before or after the controlled phase flip gate. Also note that, while we focus on states with coherence between Fock states, our results hold as well for true mixtures of photon numbers e.g. achieved by using a phase-randomized laser, beam splitters, and wave plates to impart relative phase shifts on different time bins \cite{vanEnk2001}. 

To account for physical imperfections for our protool presented in the main text we derive a model
inspired by the cavity QED implementation as modeled in Ref.~\cite{Omlor2025}.
The resulting gate model corresponds to
\begin{align}
\label{eq:CPHASE_with_errors}
  \ket{k}_q \ket{l}_0 & \to  (-1)^{k \, l} \ket{k}_q \left( \sqrt{\eta} a_{u_0}^{\dag} + \sqrt{1 - \eta} a_{u_0,L}^{\dag} \right)^l \ket{\emptyset} , \\
  \ket{k}_q \ket{1}_{\perp} & \to \ket{k}_q (\sqrt{\eta} a_{u_{\perp}}^{\dag} + \sqrt{1-\eta} a_{u_{\perp},L}^{\dag}) \ket{\emptyset} .
\end{align}
Here $\eta$ describes the efficiency of the process, where the reduced state of the qubit-channel-photon system can be found by tracing over the loss channels $a_{m,L}$ (with $m=u_0,u_{\perp}$).
We can also view the parameter $\epsilon$ to encode the (combined) infidelity of the single photon gate such that 
to achieve a good gate $\epsilon \ll 1$ is needed, hence justifying the perturbative treatment.
For simplicity, we disregard pulse distortion as it should also be small for good gates.
Lastly, the two photon component experiences no phase change according to Eq.~\eqref{eq:CPHASE_with_errors}. We emphasize that, upon detection of a single photon, the terms corresponding to vacuum disappear from the final state, and we are left only with multi-photon components and infidelities due to population of the mode $u_\perp$. Thus when the multi-photon populations are negligible and the population of $u_\perp$ is small, we approximately recover the ideal controlled phase flip gate in Eq. \eqref{eq:CPHASEqx}.

\subsection{Reflection based RSP with Intensity Encoding}\label{app:intensityRSP}
In the main text and above we focused on discussing the R-RSP approach using a conditional phase flip.
Here we discuss how one can use an intensity encoding instead \cite{nemoto-2014-photon_archit_scalab_quant_infor_proces_diamon,Omlor2025},
where if the qubit is in state $\ket{0}$ the incoming pulse is reflected, while if the state is $\ket{1}$ loss is induced and the (one-photon) component ideally is lost completely.
For reflection based entanglement generation the intensity encoding has been shown to have favorable protection from finite bandwidth effects of the single photon pulse \cite{Omlor2025}.

Ideally, intensity encoding corresponds to transfer functions $r_k \to 1-k$ ($k=0,1$).
Considering the client sends the single photon state $(\ket{E} + e^{i\phi} \ket{L})/\sqrt{2}$, with early (late) time-bin $\ket{E}$ ($\ket{L}$) and relative phase $\phi$.
The server prepares the receiving qubit in state $\ket{+} = (\ket{0} + \ket{1})/\sqrt{2}$, and the qubit first interacts with time-bin $\ket{E}$, then a NOT gate is applied on the qubit, followed by interaction with the second time-bin $\ket{L}$.
After the full interaction, the non-normalized state in the single-photon subspace within the heralding channel is
\begin{align}
\frac{\sqrt{\eta_t \eta_1}}{\sqrt{2}} \frac{1}{\sqrt{2}} \left[ e^{i \phi} \ket{L} \ket{0}_q + \ket{E} \ket{1}_q \right] ,
\end{align}
where $\eta_t$ ($\eta_1=\tilde{\eta}_0$) is the transmission (interaction) efficiency.
Removal of the time-bin information followed by projection on the single-photon subspace can be
described by a measurement with projections $(\ket{E} \pm \ket{L})/\sqrt{2}$.
This measurement can be implemented using a photon-number resolving measurement after interfering the time-bins with a 50:50 BS after delaying the first time-bin.
The success probability is ${\eta_t \eta_1 \eta_d}/{2}$, which includes the detector efficiency $\eta_d$ and is fundamentally limited by $1/2$ due to the intrinsically non-unitary qubit-photon interaction.
Furthermore, we note that both time-bins experience the same interaction.

\section{R-RSP in detail with imperfections}\label{app:imperfections}
In the main text we discussed the idealized protocol, and presented the central results in presence of imperfections.
Here we provide a model to calculate the success probability and a bound or approximation for the fidelity in the presence of imperfections.

\subsection{Photon mode error}
We begin our discussion with an error introduced in the single photon protocol, if part of the photon does not adhere to the gate and provide further details for using a linear optics model including various decay channels in the next section.
For now we consider that each time-bin's annihilation operator can be composed as
\(a_p \to (\sqrt{1 - \epsilon} a_p + \sqrt{\epsilon} a_{p,\perp})\).
The client state then becomes
\begin{align}
  \label{eq:single-photon-fid}
  \ket{\psi_c} \to \hspace{-3mm} \sum_{\vec{x} \in \{0,1\}^n} \hspace{-3mm} \sqrt{\eta_x} c_x e^{i \phi_x} (\sqrt{1 - \epsilon} a_x^{\dag} + \sqrt{\epsilon} a_{x,\perp}^{\dag}) \ket{\emptyset} .
\end{align}
As a lower bound for the fidelity, we can assign $0$ fidelity to the part $\propto \sqrt{\epsilon} a_{p,\perp}$ and take the success probability to be unaffected by $\epsilon$.
We then find the single-photon fidelity bound
\begin{align}
    \label{eq:fidelity_single_photon_SM}
    F_s \ge 1 - \epsilon .
\end{align}

\subsection{Two-photon error and photon losses}
Having discussed a first type of imperfection, we now discuss the effect if the client emits more than a single photon, as well as losses.
Because the ideal spin-photon interaction is diagonal in the computational basis, we focus only on the photonic wavefunction and emphasize that the state of the qubit register is given by \(\mathbf{H}^{\otimes n} U_x^m \ket{+}^{\otimes n}\) conditioned on the presence of \(m\) photons in time-bin \(x\).
Here, for simplicity we assume that the photon pulse is sufficiently long that the two-photon component (approximatly) performs the square of the interaction of the single photon, since the spectrum needs to be sufficiently narrow to achieve a good fidelity.

The ideal (lossless) linear optics transformation used by the client to prepare a single photon distributed over the time-bins with different phases is
$a^{\dag} \to \sum_x c_x e^{i \phi_x} a_x^{\dag}$.
We also use linear optics to model photon losses at different stages and thus map
$a_x^{\dag} \to \sqrt{\eta_x} a_x^{\dag} + \sum_{l=0}^n \sqrt{L_{x,l}} a_{x,l,L}^{\dag}$,
with the total efficiency for a photon in time-bin $x$ given by $\eta_x = \eta_t \eta_d \prod_{k=0}^{n-1} \eta_{x_k} = \eta_t \eta_d \eta_0^{n-H(x)} \eta_1^{H(x)}$ as discussed in the main text.
Additionally, we use the probability to loose the photon in time-bin \(x\) before the $q$-th qubit interaction 
$L_{x,0} = 1 - \eta_t$ ($q=0$),
$L_{x,1} = (1-\eta_{x_0}) \eta_t$ ($q=1$), and
$L_{x,q} = (1-\eta_{x_{q-1}}) \eta_t \prod_{k=0}^{q-2} \eta_{p_k}$ ($1<q<n$) as well as the probability to loose the photon after interacting with the last qubit
$L_{x,n} = (1-\eta_{x_{n-1}} \eta_d) \eta_t \prod_{k=0}^{n-2} \eta_{x_k}$.
Corresponding to the loss probabilities we also introduce the annihilation operators of the loss channels \(a_{x,l,L}\).
Here we use  the propagation efficiency $\eta_1$ ($\eta_0$) with(out) cavity interaction and where $x_k$ represents the $k$'th digit of the binary representation of $\vec{x}$.

Having discussed our model for photon losses using linear optics, we can now account for up to two photons in the client state.
To this end we use $[\alpha_0 + \alpha_1 a^{\dag}+ \frac{\alpha_2}{\sqrt{2}} (a^{\dag})^2 ]\ket{\emptyset}$ with $\alpha_m$ the amplitude of the $m$-photon component.
Applications of the above model for the linear optics, transmission and interaction with the qubits including losses, yielding a full state

\begin{widetext}
\begin{align}
     \ket{\psi_c} \to \Bigg\{\alpha_0 +  \sum_x & \left[ c_x e^{i \phi_x} (\sqrt{\eta_x} a_x^{\dag} + \sum_{l=0}^n \sqrt{L_{x,l}} a_{x,l,L}^{\dag}) \right] \\
     & \left[ \alpha_1 + \frac{\alpha_2}{\sqrt{2}} \sum_{x'} c_{x'} e^{i \phi_{x'}} (\sqrt{\eta_{x'}} a_{x'}^{\dag} + \sum_{l'=0}^n \sqrt{L_{x',l'}} a_{x',l',L}^{\dag}) \right]  \Bigg\} \ket{\emptyset} . \notag
\end{align}
If the client uses a single photon source we have $\alpha_m = \delta_{0,m}$, 
while for a weakly pulsed laser with amplitdue $\alpha_1 = \alpha$ we have 
$\alpha_0 \approx 1 - \abssq{\alpha}/2$,
$\alpha_1 \approx \alpha$,
and
$\alpha_2 \approx \alpha^2/\sqrt{2}$.

The part of the wavefunction corresponding to the ideal single-photon interaction with the qubit register is
\begin{align}
\label{eq:single-photon-non-norm-state}
     \ket{\chi} = \sum_x c_x e^{i \phi_x} \sqrt{\eta_x} a_x^{\dag} \left[ \alpha_1 + \sqrt{2} \alpha_2 \sum_{x'} c_{x'} e^{i \phi_{x'}} \sqrt{L_{x',0}} a_{x',0,L}^{\dag} \right] \ket{\emptyset},
\end{align}
and corresponds to the part of the state where no photon of the single photon term \(\propto \alpha_1\) is lost, or a single photon of the two photon term \(\propto \alpha_2\) is lost before interaction with the first qubit.

When using a photon number resolving detector the part of the photon state that contributes to the success probability are all states that have only one photon shared between the modes \(a_x\).
This includes the additional non-ideal part of the wave function
\begin{align}
\label{eq:two-photon-partial}
     \ket{\chi_\perp} = \sqrt{2} \alpha_2 \sum_x c_x e^{i \phi_x} \sqrt{\eta_x} a_x^{\dag} \left[ \sum_{x'} \sum_{l=1}^n c_{x'} e^{i \phi_{x'}} \sqrt{L_{x',l}} a_{x',l,L}^{\dag} \right] \ket{\emptyset} ,
\end{align}
where two photons arrive at the server, but one is lost during the interaction with the register.
Combining these we calculate the success probability
\begin{align}
    \label{eq:heralding-probability}
    P & = \braket{ \chi | \chi } + \braket{ \chi_{\perp} | \chi_{\perp} } = \abssq{\alpha_1} \sum_x \eta_x \abssq{c_x}
    + 2 \abssq{\alpha_2} \sum_{x,x'} \abssq{c_x} \abssq{c_{x'}}  \eta_x (1 - \eta_{x'}) \\
    & = \abssq{\alpha_1} P_0 + 2 \abssq{\alpha_2} P_0 \left( 1 - P_0 \right) , \notag
\end{align} 
where we also used the requirement of the main text that the probability for a click to stem from a time-bin is balanced, which corresponds to the condition \(\abssq{c_x} \propto \eta_x^{-1}\) (and \(\sum_x \abssq{c_x} = 1\)).
With this we find
\begin{align}
\label{eq:sum_cp_etap}
P_0 = \sum_{x} \abssq{c_x} \eta_x = \frac{2^n}{\sum_{x=0}^{2^n - 1} \eta_x^{-1}} = \frac{2^n \eta_t \eta_d}{\sum_{x=0}^{2^n - 1} \eta_0^{H(x)-n} \eta_1^{-H(x)}}  
= \eta_t \eta_d \left( \frac{2 \eta_0  \eta_1}{\eta_1 + \eta_0} \right)^n 
\le \eta_t \eta_d [\min(\eta_0, \eta_1)]^n \le 1 ,
\end{align}
in agreement with the expression given in the main text.
Herein we used combinatorics to calculate
    $\sum_{x=0}^{2^n - 1} (\eta_0/\eta_1)^{H(x)} = \sum_{k=0}^{n} ({n!}/{k! (n-k)!}) (\eta_0/\eta_1)^{k} = (1 + \eta_0/\eta_1)^n$.
The second equality uses the observation that, in all possible $n$-digit bit strings, there are $n! / k!(n-k)!$ bit strings containing exactly $k$ ones.

Analogously, we use Eq.~\eqref{eq:single-photon-non-norm-state} to calculate the probability that a single click heralds the ideal single-photon interaction, this yields
\begin{align}
  \label{eq:single-photon}
  P_S = \frac{\braket{ \chi | \chi }}{P}
  \approx 1 - \frac{2 \abssq{\alpha_2}}{\abssq{\alpha_1}} \left( \eta_t - P_0 \right)
  \approx 1 - \frac{2 \abssq{\alpha_2}}{\abssq{\alpha_1}} P_0 \left[ \frac{1}{\eta_d} \left( \frac{\eta_1 + \eta_0}{2 \eta_0 \eta_1} \right)^n - 1 \right] .
\end{align}
We can use this expression to get a simple expression to estimate (or lower bound) the fidelity, by using
$P_S F_S$ where $F_S$ is the fidelity achieved in the single photon case.

We emphasize that accounting for the fact that each part of the wavefunction $\ket{\phi}$ leads to a fidelity contribution $F_k = \abssq{\braket{\phi | +_{\vec{\theta}}}} \ge 0$, we can use $P_S$ as a lower bound for the fidelity in the presence of photon losses.
If we additionally account for the single photon error model above we find the combined lower bound $F \ge (1-\epsilon) P_s$.

To estimate the fidelity one can extend upon this bound by calculating the fidelity contribution of $\ket{\chi_\perp}$ instead of lower bounding it with $F_\perp \ge 0$.
For a single qubit $n=1$, this is largely simplified as there are only two-time bins. As both time-bins are generated from the same source using linear optics, and we are interested in losses after transmission in $\ket{\chi_\perp}$, we find two contributions to the density matrix:
if a photon from the early time-bin is lost we have the correct interaction with the time-bin and herald the correct phase
while if a late (interacting) photon is lost there is a phase error.
Averaging over all possible phase errors corresponds to $\frac{1}{2 \pi} \int_0^{2 \pi} \cos^2 \phi d\phi = \frac{1}{2}$,
such that we assign the fidelity contribution $\frac{\eta_0 + \eta_1/2}{\eta_0 + \eta_1} \ge 1/2$ to $\ket{\chi_\perp}$.
Resulting in the average fidelity estimate for $n=1$:
$F = P_S + (1-P_S) \frac{\eta_0 + \eta_1/2}{\eta_0 + \eta_1} = 1 - \frac{1}{2} \frac{\eta_1}{\eta_0 + \eta_1} (1 - P_S)$.
Inserting $P_S$ for $n=1$ leads to $F \approx 1 - \frac{2 \abssq{\alpha_2}}{\abssq{\alpha_1}} P_0 \frac{1}{4} \left[ \frac{1}{\eta_0 \eta_d} - \frac{2 \eta_1}{\eta_0 + \eta_1} \right]$.
For a coherent state this leads to the rate-fidelity tradeoff of Eq.~\eqref{eq:rate-fidelity-n=1} of the main text.

We can extend this argument to $n$ qubits to get a fidelity estimate instead of a lower bound using a zero-fidelity contribution.
To this end we assign to the different terms of $\ket{\chi_{\perp}}$ a fidelity dependent on the time-bin of the lost photon and when the photon was lost.
Both of those are encoded in loss probabilities $L_{x',l}$.
Losing a photon from time-bin $x'$ before interacting with qubit $l$ will lead to a wrong phase imprinted on $\sum_{m=0}^{l-1} x_m'$ (the Hamming weight of the first $l$ binary-digits of $x'$) many qubits.
Averaging over the random phases (analogous to the above case for $n=1$) we find the fidelity contribution $F_{x',l} = 2^{-\sum_{m=0}^{l-1} x_m'}$.
The combined contribution is
$\frac{2 \abssq{\alpha_2} P_0}{P} \sum_{x = \{0,1\}^n} \sum_{l=1}^n \abssq{c_x} L_{x,l} F_{x,l}$ which is straight forward to evaluate numerically for a given set of parameters. 

Lastly, while we did not do this for concreteness and simplicity of the analytical analysis, we note that the preceding calculations can be further extended to account for non-photon number resolving detectors by accounting for the term
\begin{align}
\label{eq:two-photon-trans}
     \ket{\chi_2} = \frac{\alpha_2}{\sqrt{2}} \sum_{x,x'} c_x c_{x'} e^{i (\phi_x + \phi_{x'})} \sqrt{\eta_x \eta_{x'}} a_x^{\dag} a_{x'}^{\dag} \ket{\emptyset} .
\end{align}
This term leads to wrong phases, which is straightforward to calculate in the limit of narrow-band pulses; the weights can be read off from Eq.~\eqref{eq:two-photon-trans}.
The detailed effect of missing photon number resolution depends on the way the time-bin information is removed e.g., if two independent ports of the QFT click then the photon number is known to be bigger than one. From our model it is easy to estimate a worst case fidelity bound by substituting \(P \to P+\braket{ \chi_2 | \chi_2 }\) into the above calculation.

\end{widetext}

\subsection{Quantum Fourier transform to remove time-bin information}\label{app:QFT}

In the main text, we assumed for simplicity, that the time-bins become indistinguishable during detection. While theoretically it is possible to create a perfect frequency-resolving single photon detector \cite{ProppPOVM} to perfectly remove time-bin information, in practice it is simpler to use a time lens \cite{Donohue2016} or optically-implemented quantum fourier transform (QFT) \cite{Barak2007} in conjunction with a time-resolving single photon detector so that detection of a single photon at a particular heralding port projects onto the final many qubit state, up to local phase corrections. Here, we go through how this can effectively be achieved using a QFT where each output mode is routed to a single photon detector (either one for each output port, or a single detector multiplexed in time).
Because this section only demonstrates the feasibility of removing the time-bin information, we only account for the unitary evolution and disregard losses for simplicity.

First, the time-bins are routed such that, after interacting with the qubit register, they are the inputs of an optical QFT and the outputs of the QFT terminate in detectors.
Heralding on detection of a single photon, this projects onto a Fourier state $\ket{k} = \sum_{x=0}^{2^n-1} \exp(2 \pi i k x / 2^n) / \sqrt{2^n} \ket{1}_p$ with \(k=0,\dots,2^n-1\).
After the interaction between the time-bins and the qubit register [see Eq.~\eqref{eq:CPHASE}], projection on $\ket{k}$ leads to
\begin{align}
  \label{eq:QFT}
  \bra{k} \mathbf{H}^{\otimes n} \mathbf{U} \ket{\psi_c} \ket{+}^{\otimes n}  
  = \sum_{x=0}^{2^{n-1}} \frac{e^{2 \pi i k x / 2^n}}{\sqrt{2^n}} c_x e^{i \phi_x} \ket{\vec{x}},
\end{align}
which becomes the target state \eqref{eq:targetstate} for the right choice of $c_x$ and $\phi_x$ after applying a $Z$ rotation of an angle $- 2 \pi k 2^q / 2^n$ on the $q$th qubit. 
This proves that depending on the outcome of the heralding port $k$ after the Fourier transform one can correct the state by a known set of single qubit gates.


\section{Security}

In terms of single-qubit RSP, we have seen that our R-RSP shares a similar performance to SC RSP, but without the need for phase stabilization. This is more than an engineering convenience, but a fundamental improvement in terms of security for protocols with weak coherent pulses \cite{Diamanti2016}; the security proofs of existing RSP schemes with low-intensity pulsed lasers rely on the phase randomization assumption to generate a statistical mixture of Fock states \cite{Dunjko2012}. Indeed, the security proof given in Ref. \cite{Garnier2024} for DC-RSP with weak coherent pulses can be used \emph{verbatim} for our R-RSP protocol for $n=1$ qubit RSP. Emphatically, the state emitted by the client is \emph{precisely} the same as for DC-RSP, and thus inherits identical security in the abstract cryptography framework \cite{Kapourniotis2024}. It is important to note that the client must be able to vary the relative phases between pulses without disturbing the time bins or otherwise revealing the values of the phases to any malicious party; otherwise, all security is lost. 

Generalizing security to $n>1$ qubits requires adapting the proof given in Ref.~\cite{Garnier2024} to qudits, and is left here as an open challenge to our colleagues. Nonetheless, with true single photon sources our protocol is secure for arbitrary $n$, as is the case for $n=1$ with single photon sources; it reduces to photonic quantum state teleportation with a qudit \cite{Luo2019,Hu2020}, with privacy guaranteed by the no-cloning theorem \cite{Wootters1982}. Indeed, one can go further and imagine reversing our R-RSP protocol into a measurement-only many-qubit RSP protocol, as we detail in the next section. This reversed R-RSP protocol (R$^3$SP) leaves the client only measuring quantum states from the server, in line with the original BQC proposal \cite{Broadbent2009} and subsequent implementations \cite{Hayashi2015,Drmota2024}.

Our approach can also be extended to multi-user BQC applications. For example, in Ref.~\cite{Polacchi2023} many users apply single-qubit rotations to a shared propagating photon, enabling quantum secured many-client computation. It is natural to consider an extension to their protocol where, using R-RSP, many qubits are modified by many users by manipulating the phases imparted on the temporal modes in the single photon superposition.


\section{Reversed R-RSP (R$^3$SP)}\label{app:R-R-RSP}

In the main text, we have considered R-RSP with an emission based client; however, it is also straigthforward to implement R-RSP in reverse. This achieves a measurement-only-like protocol where detection of a single photon at the side of the client, after the client applies phase rotations, heralds an $n$-qubit state on the server.

While more demanding on the client-side, it is a simpler system to analyze the security of; the client never sends their angles to the server, and instead relies on matter-light entanglement to teleport the information. However, there is again no free lunch; the cost is that the server must now hold the full qubit register during communication time, and attempts may not be made continuously, which could significantly lower the communication rate. 

We consider such a protocol in the idealized pulse-matched and single-photon source implementation; the generalizations to weak coherent pulses and imperfections are analogous to what we have already discussed. The server begins by preparing $n$ qubits in the state $\ket{+}^{\otimes n}$ and generates a single photon in the $2^n$-mode superposition state $\sum_{\vec{x}} c_x \ket{1}_x$, with the coefficients $c_x$ chosen such that $ \abs{c_x}^{-2} = \eta_x {\sum_{\vec{y} \in \{0,1\}^n} \eta_y^{-1}}$ as in R-RSP, with $\eta_x$ the total loss experienced by that mode. 

Consider generating the resource state in Eq. \eqref{eq:resourcestate}, and then sending the photonic part to the client. 
The client now imparts phases $e^{i\phi_x}$ on each mode $x$, and uses a time lens or QFT to erase the which-time-bin information. Now, detection of a single photon occurs with probability $P_0$ as calculated in Eq. \eqref{eq:detectionprobfinal} and heralds a state

\bea\label{eq:RRRSPstate2}
\frac{1}{\sqrt{2^n}} \sum_{\vec{x}} e^{i\phi_x} \ket{\vec{x}}.
\eea 
The phases are chosen to satisfy $\phi_x = \sum_{l=0}^{n-1} \theta_l x_l$ as in the main text, such that the result is the $n$-qubit RSP state in Eq. \ref{eq:targetstate}. 

\section{Numerical Simulation}\label{app:numerics}
In the main text we presented the rate of remote state preparation for a qubit register of length $n=8$ as a function of distance if all successes need to occur within a window of $2000/2^{k-1}$ trials.
The code generating and plotting the data is available as Ref.~\cite{CODE}.
In there we use that for $k=n$ we can analytically estimate the rate, and we calculate the rate numerically for the remaining pairs of $k q = n$ using a Monte Carlo approach by averaging over 1000 trajectories.
The parameters are inspired from silicon vacancy centers in diamond \cite{nguyen-2019-quant_networ_nodes_based_diamon,nguyen-2019-integ_nanop_quant_regis_based},
where 
a coherence time exceeding $1\,$ms and 
a cooperativity $C=38$ have been achieved while coupling to time-bins with a separation parameter of $T_{\text{TB}} = 30\,$ns.
Furthermore,
we assume 
the attenuation length of the fiber $L_{\text{att}} = 10 / (0.2 \ln 10)\,\text{km}\approx 21.7$\,km,
such that we have 
the transmission losses $\eta_t = exp(-L/L_{\text{att}}) \eta_{t,\text{intrinsic}}$.
For simplicity we assume the remaining intrinsic efficencies and the detection efficency are $\eta_{t,\text{intrinsic}} = \eta_0 = \eta_d = 0.9$.

\begin{figure}[t]
\centering
\includegraphics[width=\linewidth]{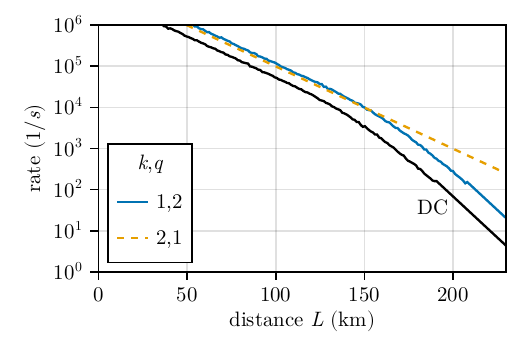}
\caption{\label{fig:2qubits}
Rate to prepare $n=k q =2$ qubits within $2000/2^{k-1}$ attempts using single-photon sources as a function of distance.
Different distribution in $q$ batches (photon detections) and $k$ qubits prepared using a single photon are encoded in the linestyle, see legend.
The remaining parameters are the same as for Fig.~3 of the main text.
}
\end{figure}

In the main text we noted that for larger $n$ we expect that preparing multiple qubits with a single photon ($k>1$) becomes favorable for shorter distances.
In Fig.~\ref{fig:2qubits}, we display the same result as in Fig.~3 of the main text but for $n=2$.
Comparing to Fig.~3 where $n=8$, the regime where the $k>1$ protocols dominates over the DC protocol occurs at a larger distance.
Verifying the general expectation and that for given fixed window size, the distance at which asymptotic scaling $R\sim\eta_t^q$ is reached decreases as the number of RSP qubits $n$ increases.

\section{Other Protocols to Remotely Prepare Many Qubits with a Single Photon}\label{app:otherproto}

Another approach to many qubit RSP with a single photon is to utilize extra qubits to encode logical qubits in a (physical qubit) fixed-Hamming weight subspace, directly generalizing the DC-RSP protocol to many-qubit RSP. One can also probabilistically and retroactively project the qubit register generated by iterated SC RSP into a fixed Hamming weight subspace via Hamming weight projection \cite{Rethinasamy2024}, generalizing double-single click (which similarly contains a probabilistic parity check) \cite{vanDam2025}. Since both DC and double-single click only require relative phase stability, they are straightforwardly amenable to security analyses \cite{Kapourniotis2024,Garnier2024}.

For these approaches, a different light-matter interaction is utilized: conditional photon emission of a single qubit  \cite{Barrett2005}, where if the qubit state is $\ket{0}$ no light is emitted, and if the qubit state is $\ket{1}$ a single photon is emitted by the server qubit. Thus such a register, prepared in a Hamming weight $1$ subspace i.e. a W state, produces a single photon in superposition over all modes. This can be combined on a beam splitter with a photon from the client to remotely prepare $n$ physical qubits on the server in a state 

 \bea\label{lightmatterstate}
\ket{\psi_s} = \frac{1}{\sqrt{n}} \sum\limits_{\vec{x}:|\vec{x}|=1} e^{i\sum_i\phi_i x_i} \ket{\vec{x}}.
\eea This protocol natively produces phase-rotated W-states with no overhead in terms of memory qubits (as these have fixed Hamming weight). It can also be used to generate GHZ states with minimal overhead for even-numbered qubit registries: by applying bit flips to half the qubits in a GHZ state $\ket{\rm GHZ} = \ket{\vec{0}\vec{0}} +  \ket{\vec{1}\vec{1}} \rightarrow \ket{\vec{0}\vec{1}} +  \ket{\vec{1}\vec{0}}$, we encode the GHZ state in the form of a generalized-$W$ state i.e. Eq. \eqref{lightmatterstate} but with the Hamming weight $|\vec{x}|$ equal to half of the qubit register size. 

Emphatically, protocols of this form do not efficiently produce phase-rotated many qubit states that can be produced with a single transmitted photon from the client; the memory overhead is exponential. Phase-rotated graph states and product states combine the full span of Hamming weights (like GHZ states) and the maximal number of independent phases (like W states), and only admit an exponential fixed Hamming weight encoding. This challenge is due to the link between Hamming weight and photon number in single qubit conditional emission. With a full quantum comptuer one can break this link by adding a single auxiliary qubit and permuting between $2^n$ $n$-qubit Toffoli gates acting on it; in each permutation, bitflips are used to select a partifcular branch $\ket{\vec{x}}$ that will entangle with the field, and then the Toffoli is applied again to restore the auxiliary qubit. With this, it is possible to generate our light-matter resource state for many-qubit RSP: $\frac{1}{\sqrt{2}^n}\sum_{\vec{x}}\hat{a}_x\vac\ket{\vec{x}}$. This can then be measured by the client as in R$^3$SP, or interfered on a $2^{n+1}$ port interferometer with a photon emitted by the client generalizing DC-RSP. However, this approach is not straightforward to implement, as it involves performing exponentially many $n$-qubit gates. 

The difficulties associated with using the single qubit conditional emission motivate our development of the engineered multi-qubit light-matter interaction for our R-RSP protocol; this natively decouples photon number, Hamming weight, and qubit number. While an exponential time bin overhead is not ideal, it is much better than an exponential overhead in the number of physical qubits or the number of multi-qubit gates, enabling the comparably efficient single-photon encoding of many qubits utilized in this work.

\section{Many-Qubit RSP: Asymptotic Form}\label{app:asymptot}

In the limit where the probability of successfully preparing $n$ RSP qubits in $q$ batches of $k$ qubits within a window of $w$ attempts approaches zero (e.g. because $\eta_t \rightarrow 0$), the rate of successful $n$-qubit RSP can be approximated straightforwardly using the results of Ref. \cite{Davies2024}. There, as the probability of an attempt succeeding $p$ becomes small, it is shown that one must wait $\sim p^{-q} {w-1 \choose q-1}^{-1}$ trials to get $q$ successes within a window of $w$ trials. To calculate a rate with this for our problem, we must also account for each trial taking a time that depends on the number of time-bins $\sim T_{\rm TB} 2^{k-1}$. We neglect final communication time, as classical corrections can co-propagate with the R-RSP so that communication time contributes only to a latency and not an infidelity. We also must scale the effective window size $w\rightarrow w_0 2^{-{k-1}} $ to prepare the same number of qubits in the same lab-time. Thus, for RSP of $n$ qubits in $q$ batches of $k$ qubits, we arrive at an asymptotic rate 

\begin{widetext}
\bea\label{eq:rateapproxrepeat}
\lim_{\eta_t \rightarrow 0} R_q  =\eta_t^{q}\eta_d^{q} 2^{-{k-1}} \left( \frac{2 \eta_0 \eta_1}{\eta_0 + \eta_1} \right)^n {w_0 2^{-k+1} - 1 \choose q -1} T_{TB}^{-1}.
\eea     
\end{widetext}

As mentioned in the main-text, many-qubit RSP makes use of coherence between many temporally multiplexed pulses and is a form of quantum multiplexing, introduced in Ref. \cite{LoPiparo2019} and defined in general in Ref. \cite{Propp2025}. As in Ref. \cite{Propp2025}, we can calculate the multiplexing advantage for an $n$-qubit RSP in the low-transmission limit

\bea\label{eq:multiplexing}
m_n^{(q)} = \frac{R_q}{R_n}  &\stackrel{\eta_t\rightarrow 0}{=}& \eta_t^{q-n}\eta_d^{q-n}2^{n-k} \frac{{w_0 2^{-k+1} - 1 \choose q -1} }{{w_0  - 1 \choose n -1} }.\nonumber \\ 
\eea 

In comparison to the other beyond-classical multiplexing strategies discussed in the literature, our technique leads to a divergent multiplexing advantage in the small-transmission limit; this is due to the reduction in the number of successes needed e.g. the number of photons that need to be transmitted ($q$) from client to server in order to remotely prepare $n$ qubits.

\section{Proof that $2^n$ temporal modes are required to encode $n$ qubits in a single photon}\label{app:proof}

Another notable feature of our R-RSP protocol is, that while it requires exponentially large number of time bins when preparing the full register with a single photon, no additional memory qubits are needed to implement $n$-qubit RSP with a single photon, even implemented with weak coherent pulses (WCPs). 
We now show why we need $2^n$ pulses to encode $n$ phases into $n$ qubits using a single photon interacting with an $n$-qubit register.

  We begin with a formal proof of the requirement, arguing from Hilbert space dimension. We will only consider RSP of equatorial qubit states $\ket{+_\theta}=\frac{1}{\sqrt{2}}(\ket{0}+e^{i\theta}\ket{1})$, as these are the only states strictly required for verifiable blind quantum computation \cite{Fitzsimons2017}. 

  \begin{proof}
    Assume for contradiction that there exists a deterministic, unit-fidelity encoding of every $n$-qubit product equatorial state $\ket{\psi(\vec{\theta})}=\frac{1}{\sqrt{n}} \sum\limits_{\vec{x}} e^{i\sum_i\theta_i x_i} \ket{\vec{x}}$ into a single photon in superposition over $M$ temporal modes with $M<2^n$. Thus, the single photon lives in a Hilbert space $\mathcal{H}_\gamma$ with dimension $D=M$.

    Let $V={\rm span}\left\{\,\ket{\psi(\vec{\theta})}: \vec{\theta}\in[0,2\pi)^n\,\right\} = \left(\mathbb{C}^{2}\right)^{\otimes n}$.

    Let $I$ be a linear isometry encoding $n$ equatorial qubits in the single photon $I:V\rightarrow \mathcal{H}_\gamma$. By definition, an isometry is injective so that ${\rm dim}(V) \leq {\rm dim}(H) \rightarrow 2^n \leq M$. 

    We thus arrive at our contradiction: either $M\geq2^n$, or $I$ cannot be an isometry. And if $I$ is not an isometry, it cannot deterministically and faithfully encode the qubit states \cite{Chefles2000}.  

\end{proof}


\end{document}